\documentclass{JHEP3}

\usepackage{amsfonts,amsmath}				
\usepackage{longtable}						
\usepackage{enumerate}						
\usepackage{pstricks,pst-node}				

\newcommand{\g}{\mathfrak{g}}				
\newcommand{\s}{\mathfrak{s}}				
\newcommand{\R}{\mathbb{R}}					

\newcommand{\mline}[2]						
	{$\begin{array}{c} #1 \times \\ \times #2 \end{array}$}	

\newcommand{\simplie}[2]					
	{{Class \tt \href{http://strings.fmns.rug.nl/SimpLie/javadoc/edu/rug/hep/simplie/#1/#2.html}{#2}}}

\newcommand{\refSL}[1]						
	{[\ref{sl:sl}\ref{#1}]}

\psset{unit=.65cm,linewidth=.5pt,radius=.2}	

\title{$E_{11}$ and the Embedding Tensor}

\author{Eric A. Bergshoeff, Teake A. Nutma\\
	Centre for Theoretical Physics, University of Groningen, \\
    Nijenborgh 4, 9747 AG Groningen, The Netherlands \\
    \email{E.A.Bergshoeff@rug.nl, T.A.Nutma@rug.nl }}
\author{Iwein De Baetselier\\
	Instituut voor Theoretische Fysica, Katholieke Universiteit Leuven, \\
    Celestijnenlaan 200D, B--3001 Leuven, Belgium}

\abstract{
We show how, using different decompositions of $E_{11}$,
one can calculate the representations under the
duality group of  the so--called ``de-form'' potentials.
Evidence is presented that  these potentials are in
one-to-one correspondence to the embedding tensors
that classify the gaugings of all maximal gauged supergravities.
We supply the computer program underlying our calculations.
}

\preprint{UG-07-03\\KUL-TF-07/10}

\begin{document}

\section{Introduction}

For some time now it has been conjectured that the infinite--dimensional
algebra of $E_{11}$ is the underlying symmetry of eleven--dimensional
supergravity or M--theory \cite{West:2001as,Schnakenburg:2001ya,Kleinschmidt:2003mf}.
One piece of evidence for this was provided by the proof
\cite{Bergshoeff:2005ac,Bergshoeff:2006qw} that the top--forms \footnote{By a
top-form in D dimensions we mean a gauge field with D anti--symmetric indices.
Such gauge fields couple to space--filling branes such as the D9--brane in 10
dimensions. Note that, due to the gauge symmetries, top--forms are
{\it inequivalent} to (the product of a Levi--Civita tensor and) scalars.}
consistent with ten-dimensional IIA and IIB supergravity are precisely the ones
predicted by $E_{11}$ \cite{Kleinschmidt:2003mf,West:2005gu}.
The existence of these top--forms do not follow from the representation theory
of the supersymmetry algebra since they do not describe physical degrees of
freedom. Their existence can be proven by showing that the ten-dimensional
supersymmetry algebra can be realized on these fields.
A prime example of a top--form is the Ramond--Ramond 10--form that couples to
the D9--brane of Type IIB string theory. It turns out that this ten--form
is part of a quadruplet of ten-forms transforming according to the ${\bf 4}$
representation of the $SL(2,\R)$ duality group \cite{Bergshoeff:2005ac}.
The nine--branes of Type IIB string theory form a non-linear doublet that is
embedded into this quadruplet \cite{Bergshoeff:2006ic}.

Another set of fields predicted by $E_{11}$ are the so--called ``de--forms''
\footnote{By a ``de-form'' in D dimensions we mean a gauge field with
$\text{D--1}$ anti--symmetric indices. Every de--form corresponds to a
deformation of the corresponding supergravity theory with a masslike parameter.
Hence the name (we thank Diederik Roest for suggesting this terminology to us).
The masslike parameter occurs as an integration constant to the equation of
motion of the de--form.}. Like the top--forms they do not follow from the
representation theory of the supersymmetry algebra.
The prime example of a de--form is the ten--dimensional nine--form
\cite{Bergshoeff:1996ui} that is related to the masslike parameter $m$ of massive
IIA supergravity \cite{Romans:1985tz}. A priori not every deformation of a
supergravity theory with a masslike parameter corresponds to a gauged supergravity.
In particular, massive IIA supergravity cannot be obtained as the gauging of
the $\R^+$ duality group. The Ramond--Ramond 1--form cannot play the role
of the candidate gauge field since it is not invariant under the $\R^+$
scaling symmetry. It is only after a torus reduction to nine dimensions that
massive IIA supergravity becomes a nine--dimensional maximal gauged supergravity
with gauge group $\R^+$ and with the Kaluza--Klein vector as the correct
gauge field.

The class of maximal gauged supergravities has been investigated in
\cite{Nicolai:2001sv,deWit:2002vt,deWit:2003hr}. It was shown that a consistent gauging requires
that the so--called embedding tensor transforms according to a specific
representation of the duality group. Possible gaugings can be explored by
verifying which gauge groups lead to an embedding tensor in this particular
representation. The embedding tensor may be viewed as a collection of integration
constants that transforms according to a representation of the duality group.
Equivalently, the integration constants may be described by corresponding
de--form potentials. Such de--forms also occur in different decompositions of
$E_{11}$. Therefore, an important piece of evidence in favor of an underlying
$E_{11}$ symmetry would be to show that the de--forms predicted by $E_{11}$
transform precisely in the representations required by imposing a consistent
supersymmetric gauging following \cite{Nicolai:2001sv,deWit:2002vt,deWit:2003hr}.
It is the purpose of this paper to show that this is indeed the case.

In the final stages of this project we received a paper by Riccioni and West
\cite{Riccioni:2007au} that contains overlap with this work. They derive
the same de--forms and top--forms in various dimensions via a different technique
and show that the resulting de--forms are in agreement with the literature.
There is also an interesting connection with work in progress by de Wit, Nicolai,
and Samtleben\,\footnote{B. de Wit, private communication.} \cite{firenze}.

This paper is organized as follows. In section 2 we review how the physical
states of all $D\le 11$ maximal supergravities occur as representations of
regular subalgebras of $E_{11}$. In section 3 we derive which de--forms and
top--forms in maximal supergravity are predicted by  $E_{11}$.
Finally, we comment on our results in the conclusions. We have included two
appendices. Appendix \ref{app:simplie} explains how we calculated the different
decompositions of $E_{11}$ using a computer program. Appendix \ref{app:results}
contains the relevant low level results of the spectrum.

\section{Physical States}

It is well known that maximal supergravities are characterized by hidden
symmetries. This means that the $(11-n)$-dimensional supergravities exhibit a
duality symmetry group $G$ of rank $0\le n \le 11$ larger than the
$SL(n,\R)\times \R^+$ symmetry group expected to follow from the
reduction of $D=11$ supergravity over an $n$--torus.
In particular, the scalars transform non--linearly under the duality group $G$
and parameterize a coset $G/K(G)$, where $K$ is the maximal compact subgroup of
$G$. We have given $G$ and $K(G)$ for the different dimensions in table \ref{table1}
alongside the corresponding decomposition of $E_{11}$ from which $G$ follows.
Note that each $G$ is maximal non--compact.

\TABLE[ht]{
	\begin{tabular}{|*{5}{c|}}
		\hline	
		$D$	& $G$							& $K(G)$				& $\dim(G/K)$	& $E_{11}$ decomposition\\
		\hline
		\hline
		&&&& \\ [-8pt]
		11	& 1								& $1$					& 0				& \tiny\begin{pspicture}(0,0)(9,1)
\Cnode[fillstyle=solid,fillcolor=black](2,1){N11723083412}
\nput[labelsep=0.2]{-40}{N11723083412}{1}
\Cnode(0,0){N21723083412}
\nput[labelsep=0.2]{-40}{N21723083412}{2}
\Cnode(1,0){N31723083412}
\nput[labelsep=0.2]{-40}{N31723083412}{3}
\Cnode(2,0){N41723083412}
\nput[labelsep=0.2]{-40}{N41723083412}{4}
\Cnode(3,0){N51723083412}
\nput[labelsep=0.2]{-40}{N51723083412}{5}
\Cnode(4,0){N61723083412}
\nput[labelsep=0.2]{-40}{N61723083412}{6}
\Cnode(5,0){N71723083412}
\nput[labelsep=0.2]{-40}{N71723083412}{7}
\Cnode(6,0){N81723083412}
\nput[labelsep=0.2]{-40}{N81723083412}{8}
\Cnode(7,0){N91723083412}
\nput[labelsep=0.2]{-40}{N91723083412}{9}
\Cnode(8,0){N101723083412}
\nput[labelsep=0.2]{-40}{N101723083412}{10}
\Cnode(9,0){N111723083412}
\nput[labelsep=0.2]{-40}{N111723083412}{11}
\ncline{-}{N21723083412}{N31723083412}
\ncline{-}{N31723083412}{N41723083412}
\ncline{-}{N41723083412}{N51723083412}
\ncline{-}{N51723083412}{N61723083412}
\ncline{-}{N61723083412}{N71723083412}
\ncline{-}{N71723083412}{N81723083412}
\ncline{-}{N81723083412}{N91723083412}
\ncline{-}{N91723083412}{N101723083412}
\ncline{-}{N101723083412}{N111723083412}
\ncline{-}{N11723083412}{N41723083412}
\end{pspicture} ~~\\ [10pt]
		IIA	& $\R^+$						& $1$					& 1				& \begin{pspicture}(0,0)(9,1)
\Cnode[fillstyle=solid,fillcolor=black](2,1){N11213619404}
\Cnode[fillstyle=solid,fillcolor=black](0,0){N21213619404}
\Cnode(1,0){N31213619404}
\Cnode(2,0){N41213619404}
\Cnode(3,0){N51213619404}
\Cnode(4,0){N61213619404}
\Cnode(5,0){N71213619404}
\Cnode(6,0){N81213619404}
\Cnode(7,0){N91213619404}
\Cnode(8,0){N101213619404}
\Cnode(9,0){N111213619404}
\ncline{-}{N21213619404}{N31213619404}
\ncline{-}{N31213619404}{N41213619404}
\ncline{-}{N41213619404}{N51213619404}
\ncline{-}{N51213619404}{N61213619404}
\ncline{-}{N61213619404}{N71213619404}
\ncline{-}{N71213619404}{N81213619404}
\ncline{-}{N81213619404}{N91213619404}
\ncline{-}{N91213619404}{N101213619404}
\ncline{-}{N101213619404}{N111213619404}
\ncline{-}{N11213619404}{N41213619404}
\end{pspicture} ~~\\ [10pt]
		IIB	& $SL(2,\R)$					& $SO(2)$				& 2				& \begin{pspicture}(0,0)(9,1)
\Cnode(2,1){N11213619404}
\Cnode[fillstyle=solid,fillcolor=lightgray](0,0){N21213619404}
\Cnode[fillstyle=solid,fillcolor=black](1,0){N31213619404}
\Cnode(2,0){N41213619404}
\Cnode(3,0){N51213619404}
\Cnode(4,0){N61213619404}
\Cnode(5,0){N71213619404}
\Cnode(6,0){N81213619404}
\Cnode(7,0){N91213619404}
\Cnode(8,0){N101213619404}
\Cnode(9,0){N111213619404}
\ncline{-}{N21213619404}{N31213619404}
\ncline{-}{N31213619404}{N41213619404}
\ncline{-}{N41213619404}{N51213619404}
\ncline{-}{N51213619404}{N61213619404}
\ncline{-}{N61213619404}{N71213619404}
\ncline{-}{N71213619404}{N81213619404}
\ncline{-}{N81213619404}{N91213619404}
\ncline{-}{N91213619404}{N101213619404}
\ncline{-}{N101213619404}{N111213619404}
\ncline{-}{N11213619404}{N41213619404}
\end{pspicture} ~~\\ [10pt]
		9	& $SL(2,\R)\times \R^+$			& $SO(2)$				& 3				& \begin{pspicture}(0,0)(9,1)
\Cnode[fillstyle=solid,fillcolor=black](2,1){N11343807624}
\Cnode[fillstyle=solid,fillcolor=lightgray](0,0){N21343807624}
\Cnode[fillstyle=solid,fillcolor=black](1,0){N31343807624}
\Cnode(2,0){N41343807624}
\Cnode(3,0){N51343807624}
\Cnode(4,0){N61343807624}
\Cnode(5,0){N71343807624}
\Cnode(6,0){N81343807624}
\Cnode(7,0){N91343807624}
\Cnode(8,0){N101343807624}
\Cnode(9,0){N111343807624}
\ncline{-}{N21343807624}{N31343807624}
\ncline{-}{N31343807624}{N41343807624}
\ncline{-}{N41343807624}{N51343807624}
\ncline{-}{N51343807624}{N61343807624}
\ncline{-}{N61343807624}{N71343807624}
\ncline{-}{N71343807624}{N81343807624}
\ncline{-}{N81343807624}{N91343807624}
\ncline{-}{N91343807624}{N101343807624}
\ncline{-}{N101343807624}{N111343807624}
\ncline{-}{N11343807624}{N41343807624}
\end{pspicture} ~~\\ [10pt]
		8	& \mline{SL(3,\R)}{SL(2,\R)}	& \mline{SO(3)}{SO(2)}	& 7				& \begin{pspicture}(0,0)(9,1)
\Cnode[fillstyle=solid,fillcolor=lightgray](2,1){N11476457064}
\Cnode[fillstyle=solid,fillcolor=lightgray](0,0){N21476457064}
\Cnode[fillstyle=solid,fillcolor=lightgray](1,0){N31476457064}
\Cnode[fillstyle=solid,fillcolor=black](2,0){N41476457064}
\Cnode(3,0){N51476457064}
\Cnode(4,0){N61476457064}
\Cnode(5,0){N71476457064}
\Cnode(6,0){N81476457064}
\Cnode(7,0){N91476457064}
\Cnode(8,0){N101476457064}
\Cnode(9,0){N111476457064}
\ncline{-}{N21476457064}{N31476457064}
\ncline{-}{N31476457064}{N41476457064}
\ncline{-}{N41476457064}{N51476457064}
\ncline{-}{N51476457064}{N61476457064}
\ncline{-}{N61476457064}{N71476457064}
\ncline{-}{N71476457064}{N81476457064}
\ncline{-}{N81476457064}{N91476457064}
\ncline{-}{N91476457064}{N101476457064}
\ncline{-}{N101476457064}{N111476457064}
\ncline{-}{N11476457064}{N41476457064}
\end{pspicture} ~~\\ [10pt]
		7	& $SL(5,\R)$					& $SO(5)$\				& 14			& \begin{pspicture}(0,0)(9,1)
\Cnode[fillstyle=solid,fillcolor=lightgray](2,1){N11678698620}
\Cnode[fillstyle=solid,fillcolor=lightgray](0,0){N21678698620}
\Cnode[fillstyle=solid,fillcolor=lightgray](1,0){N31678698620}
\Cnode[fillstyle=solid,fillcolor=lightgray](2,0){N41678698620}
\Cnode[fillstyle=solid,fillcolor=black](3,0){N51678698620}
\Cnode(4,0){N61678698620}
\Cnode(5,0){N71678698620}
\Cnode(6,0){N81678698620}
\Cnode(7,0){N91678698620}
\Cnode(8,0){N101678698620}
\Cnode(9,0){N111678698620}
\ncline{-}{N21678698620}{N31678698620}
\ncline{-}{N31678698620}{N41678698620}
\ncline{-}{N41678698620}{N51678698620}
\ncline{-}{N51678698620}{N61678698620}
\ncline{-}{N61678698620}{N71678698620}
\ncline{-}{N71678698620}{N81678698620}
\ncline{-}{N81678698620}{N91678698620}
\ncline{-}{N91678698620}{N101678698620}
\ncline{-}{N101678698620}{N111678698620}
\ncline{-}{N11678698620}{N41678698620}
\end{pspicture} ~~\\ [10pt]
		6	& $SO(5,5)$						& \mline{SO(5)}{SO(5)}	& 25			& \begin{pspicture}(0,0)(9,1)
\Cnode[fillstyle=solid,fillcolor=lightgray](2,1){N11195574684}
\Cnode[fillstyle=solid,fillcolor=lightgray](0,0){N21195574684}
\Cnode[fillstyle=solid,fillcolor=lightgray](1,0){N31195574684}
\Cnode[fillstyle=solid,fillcolor=lightgray](2,0){N41195574684}
\Cnode[fillstyle=solid,fillcolor=lightgray](3,0){N51195574684}
\Cnode[fillstyle=solid,fillcolor=black](4,0){N61195574684}
\Cnode(5,0){N71195574684}
\Cnode(6,0){N81195574684}
\Cnode(7,0){N91195574684}
\Cnode(8,0){N101195574684}
\Cnode(9,0){N111195574684}
\ncline{-}{N21195574684}{N31195574684}
\ncline{-}{N31195574684}{N41195574684}
\ncline{-}{N41195574684}{N51195574684}
\ncline{-}{N51195574684}{N61195574684}
\ncline{-}{N61195574684}{N71195574684}
\ncline{-}{N71195574684}{N81195574684}
\ncline{-}{N81195574684}{N91195574684}
\ncline{-}{N91195574684}{N101195574684}
\ncline{-}{N101195574684}{N111195574684}
\ncline{-}{N11195574684}{N41195574684}
\end{pspicture} ~~\\ [10pt]
		5	& $E_{6(+6)}$					& $USp(8)$				& 42			& \begin{pspicture}(0,0)(9,1)
\Cnode[fillstyle=solid,fillcolor=lightgray](2,1){N11083197328}
\Cnode[fillstyle=solid,fillcolor=lightgray](0,0){N21083197328}
\Cnode[fillstyle=solid,fillcolor=lightgray](1,0){N31083197328}
\Cnode[fillstyle=solid,fillcolor=lightgray](2,0){N41083197328}
\Cnode[fillstyle=solid,fillcolor=lightgray](3,0){N51083197328}
\Cnode[fillstyle=solid,fillcolor=lightgray](4,0){N61083197328}
\Cnode[fillstyle=solid,fillcolor=black](5,0){N71083197328}
\Cnode(6,0){N81083197328}
\Cnode(7,0){N91083197328}
\Cnode(8,0){N101083197328}
\Cnode(9,0){N111083197328}
\ncline{-}{N21083197328}{N31083197328}
\ncline{-}{N31083197328}{N41083197328}
\ncline{-}{N41083197328}{N51083197328}
\ncline{-}{N51083197328}{N61083197328}
\ncline{-}{N61083197328}{N71083197328}
\ncline{-}{N71083197328}{N81083197328}
\ncline{-}{N81083197328}{N91083197328}
\ncline{-}{N91083197328}{N101083197328}
\ncline{-}{N101083197328}{N111083197328}
\ncline{-}{N11083197328}{N41083197328}
\end{pspicture} ~~\\ [10pt]
		4	& $E_{7(+7)}$					& $SU(8)$				& 70			& \begin{pspicture}(0,0)(9,1)
\Cnode[fillstyle=solid,fillcolor=lightgray](2,1){N11034243376}
\Cnode[fillstyle=solid,fillcolor=lightgray](0,0){N21034243376}
\Cnode[fillstyle=solid,fillcolor=lightgray](1,0){N31034243376}
\Cnode[fillstyle=solid,fillcolor=lightgray](2,0){N41034243376}
\Cnode[fillstyle=solid,fillcolor=lightgray](3,0){N51034243376}
\Cnode[fillstyle=solid,fillcolor=lightgray](4,0){N61034243376}
\Cnode[fillstyle=solid,fillcolor=lightgray](5,0){N71034243376}
\Cnode[fillstyle=solid,fillcolor=black](6,0){N81034243376}
\Cnode(7,0){N91034243376}
\Cnode(8,0){N101034243376}
\Cnode(9,0){N111034243376}
\ncline{-}{N21034243376}{N31034243376}
\ncline{-}{N31034243376}{N41034243376}
\ncline{-}{N41034243376}{N51034243376}
\ncline{-}{N51034243376}{N61034243376}
\ncline{-}{N61034243376}{N71034243376}
\ncline{-}{N71034243376}{N81034243376}
\ncline{-}{N81034243376}{N91034243376}
\ncline{-}{N91034243376}{N101034243376}
\ncline{-}{N101034243376}{N111034243376}
\ncline{-}{N11034243376}{N41034243376}
\end{pspicture} ~~\\ [10pt]
		3	& $E_{8(+8)}$					& $SO(16)$				& 128			& \begin{pspicture}(0,0)(9,1)
\Cnode[fillstyle=solid,fillcolor=lightgray](2,1){N11991063964}
\Cnode[fillstyle=solid,fillcolor=lightgray](0,0){N21991063964}
\Cnode[fillstyle=solid,fillcolor=lightgray](1,0){N31991063964}
\Cnode[fillstyle=solid,fillcolor=lightgray](2,0){N41991063964}
\Cnode[fillstyle=solid,fillcolor=lightgray](3,0){N51991063964}
\Cnode[fillstyle=solid,fillcolor=lightgray](4,0){N61991063964}
\Cnode[fillstyle=solid,fillcolor=lightgray](5,0){N71991063964}
\Cnode[fillstyle=solid,fillcolor=lightgray](6,0){N81991063964}
\Cnode[fillstyle=solid,fillcolor=black](7,0){N91991063964}
\Cnode(8,0){N101991063964}
\Cnode(9,0){N111991063964}
\ncline{-}{N21991063964}{N31991063964}
\ncline{-}{N31991063964}{N41991063964}
\ncline{-}{N41991063964}{N51991063964}
\ncline{-}{N51991063964}{N61991063964}
\ncline{-}{N61991063964}{N71991063964}
\ncline{-}{N71991063964}{N81991063964}
\ncline{-}{N81991063964}{N91991063964}
\ncline{-}{N91991063964}{N101991063964}
\ncline{-}{N101991063964}{N111991063964}
\ncline{-}{N11991063964}{N41991063964}
\end{pspicture} ~~\\ [10pt]
		\hline
	\end{tabular}
	\caption{
		The hidden symmetries of all $3\le D\le 11$ maximal	supergravities.
		The duality groups $G$ can be read of from the decomposition of the
		Dynkin diagram of $E_{11}$; they correspond to the grey nodes. The white
		nodes form the gravity line $A_{D-1} = SL(D)$.
		In each case the scalars parameterize the coset $G/K$ such that the
		number of scalars is equal to the dimension of the coset.
	}
	\label{table1}
}

At low levels the decomposition of $E_{11}$ with respect to different regular
subalgebras should contain the physical states of maximal supergravity in
$3\le D\le 11$ dimensions. Applying the level decomposition described in Appendix
\ref{app:ld} we indeed obtain the physical states of $D$--dimensional
maximal supergravity, see table \ref{table2}. Each supergravity field transforms
as a representation of $A_{D-1}\times G$ where $A_{D-1}$ refers to the
spacetime symmetries and $G$ is one of the duality groups given in table
\ref{table1}.

It is straightforward to search for these specific supergravity fields in the
different decompositions of $E_{11}$ using the level decomposition rules
explained in Appendix \ref{app:simplie}. To perform the relevant calculations we
developed the computer program SimpLie.

\TABLE[ht]{
	\begin{tabular}{|c|*{6}{r@{\ $\times$ }l|}}
		\hline
		$D$	& \multicolumn{2}{c|}{$g_{\mu\nu}$} & \multicolumn{2}{c|}{$p = 0$}
			& \multicolumn{2}{c|}{$p = 1$} 		& \multicolumn{2}{c|}{$p = 2$}
			& \multicolumn{2}{c|}{$p = 3$} 		& \multicolumn{2}{c|}{$(p = 4)^+$} \\
			
		\hline
		\hline
		
		11	& \bf 44 	& \bf 1					& \multicolumn{2}{c|}{}
			& \multicolumn{2}{c|}{}				& \multicolumn{2}{c|}{}
			& \bf 84 	& \bf 1					& \multicolumn{2}{c|}{} \\
			
		IIA	& \bf 35	& \bf 1					& \bf 1		& 1
			& \bf 8		& \bf 1					& \bf 28	& \bf 1
			& \bf 56	& \bf 1					& \multicolumn{2}{c|}{} \\
			
		IIB	& \bf 35	& \bf 1					& \bf 1		& 2
			& \multicolumn{2}{c|}{}				& \bf 28	& \bf 2
			& \multicolumn{2}{c|}{}				& \bf 35	& \bf 1 \\
			
		9	& \bf 27	& \bf 1					& \bf 1		& 3
			& \bf 7		& (\textbf 1 + \textbf 2)& \bf 21	& \bf 2
			& \bf 35	& \bf 1					& \multicolumn{2}{c|}{} \\
			
		8	& \bf 20	& \bf (1,1)				& \bf 1		& 7
			& \bf 6		& \bf (3,2)				& \bf 15	& \bf (3,1)
			& \bf 20	& $\frac 1 2$ \bf (1,2)	& \multicolumn{2}{c|}{} \\
			
		7	& \bf 14	& \bf 1					& \bf 1		& 14
			& \bf 5		& \bf 10 				& \bf 10	& \bf 5
			& \multicolumn{2}{c|}{}				& \multicolumn{2}{c|}{} \\
			
		6	& \bf 9		& \bf 1					& \bf 1		& 25
			& \bf 4		& \bf 16 				& \bf 6		& $\frac 1 2$ \bf 10
			& \multicolumn{2}{c|}{}				& \multicolumn{2}{c|}{} \\
			
		5	& \bf 5		& \bf 1					& \bf 1		& 42
			& \bf 3		& \bf 27 				& \multicolumn{2}{c|}{}
			& \multicolumn{2}{c|}{}				& \multicolumn{2}{c|}{} \\
			
		4	& \bf 2		& \bf 1					& \bf 1		& 70
			& \bf 2		& $\frac 1 2$ $\textbf{28}^c$	& \multicolumn{2}{c|}{}
			& \multicolumn{2}{c|}{}				& \multicolumn{2}{c|}{} \\
			
		3	& --		& \bf 1					& \bf 1		& 128
			& \multicolumn{2}{c|}{}				& \multicolumn{2}{c|}{}
			& \multicolumn{2}{c|}{}				& \multicolumn{2}{c|}{} \\
			
		\hline
	\end{tabular}
	\caption{
		The occurrence of the physical states of all $3\le D\le 11$ maximal
		supergravities in the level decomposition of $E_{11}$, which are also
		listed in Appendix \ref{app:results}. The $p$--columns indicate which
		$p$--form potentials are present. All entries apart from $p=0$ are of
		the form  ``physical d.o.f. $\times$ $G$ representation,'' where $G$
		is the duality group. For $p=0$ the entries read
		``physical d.o.f. $\times$ number of scalars.''
	}
	\label{table2}
}

All fields occur in representations of $G$ except the scalars. Dividing out by
the maximal compact subgroup of $E_{11}$ means that we restrict the spectrum
to the positive levels and therefore we keep only the axions associated to the
positive root generators of $G$ and the dilatons associated to the Cartan
generators of $G$. Together, these scalars parameterize the coset $G/K$. Note
that the curl of the 4--form potential in IIB is self--dual, we therefore count
this potential as 35 degrees of freedom. Furthermore, in $D=8$ the 3--form
potential and its dual together form a doublet $\textbf{(1,2)}$. Thus the
potential is counted as $1/2 \times \textbf{(1,2)}$ as indicated in the table.
The same applies to the 2--forms in $D=6$ and the 1--forms in $D=4$. In each
dimension the total number of states adds up to 128.

\section{De--forms and Top--forms}

We now extend the analysis of the previous section to include  the de--forms and
top--forms in our calculations. The results ensue from Appendix \ref{app:results}
and are summarized in table \ref{table3}. The top--forms are identified using the
observations made in \cite{Riccioni:2006az}.

A few remarks are in order. First of all, there are no de--forms or top--forms in
$D=11$ dimensions. Furthermore, table 3 reproduces the known de--forms and
top--forms in $D=10$ dimensions \cite{Bergshoeff:2005ac,Bergshoeff:2006qw,Bergshoeff:1996ui}.
More importantly, the de--form representations coincide precisely with the
embedding tensor calculations given in \cite{Nicolai:2001sv,deWit:2002vt,deWit:2003hr}. This is
non-trivial given the fact that the de--form calculation is based upon $E_{11}$
whereas the calculations of \cite{Nicolai:2001sv,deWit:2002vt,deWit:2003hr} involve supergravity.

It is interesting to compare the de--form calculation with some of the known
results on gauged and/or massive supergravities in dimensions $D \le 10$. As
explained in the introduction, the IIA theory has a singlet massive deformation
which is the massive supergravity of \cite{Romans:1985tz}. Note that there is a
single maximal gauged supergravity in $D=10$ \cite{Howe:1997qt} but this theory
can only be defined at the level of the equations of motion. Apparently,
$E_{11}$ does not give rise to theories without an action.

\TABLE[ht]{
	\centering
	\begin{tabular}{|c||c|c|c|c|c|c|c|c|c|}
		\hline
		$D$ & IIA & IIB & 9 & 8 & 7 & 6 & 5 & 4 & 3 \\
		\hline
		\hline
		
		de-forms
			& \bf 1		&			& \bf 2			
			& \bf 6		& \bf 15	& \bf 144		
			& \bf 351	& \bf 912	& \bf 1 \\		
			
			&			&			& \bf 3			
			& \bf 12	& \bf 40	&				
			&			&			& \bf 3875 \\	
			
		\hline
		
		top-forms
			& $2 \times \bf 1$	& \bf 2		& $2 \times \bf 2$	
			& $2 \times \bf 3$	& \bf 5 	& \bf 10			
			& \bf 27			& \bf 133	& \bf 248 \\		
			
			&					& \bf 4		& \bf 4				
			& \bf 9				& \bf 45	& \bf 126			
			& \bf 1728			& \bf 8645 	& \bf 3875 \\		
			
			&					&			&					
			& \bf 15 			& \bf 70 	& \bf 320			
			&					&			& \bf 147250 \\		
			
		\hline
	\end{tabular}
	\caption{
		$E_{11}$ predictions for de--forms and top--forms in all $3\le D\le 10$
		maximal supergravities. These are representations of the respective
		duality groups $G$ given in table \ref{table1}.
	}
	\label{table3}
}

Maximal gauged supergravities in $D=9$ have been considered in
\cite{Bergshoeff:2002nv}. As observed in \cite{Riccioni:2007au} the $D=9$
de--forms agree perfectly with the triplet and doublet deformations of
\cite{Bergshoeff:2002nv}. Both deformations are gauged supergravities with a
corresponding action. The analysis of \cite{Bergshoeff:2002nv} contains more
$D=9$ maximal gauged supergravities but they do not have an action.

Maximal gauged supergravities in $D=8$ have been studied (but not exhaustively
classified) in \cite{Bergshoeff:2003ri}. Only the ones that follow from a group
manifold reduction of $D=11$ supergravity have been given. They are contracted
and non--compact versions of the $SO(3)$ gauged supergravity of \cite{SS}. These
results have been compared with the de--form calculation
\cite{Weidner:2007wt,Riccioni:2007au}. It seems there are more maximal gauged
supergravities than the ones constructed in \cite{Bergshoeff:2003ri}. Finally,
we note that extensive studies of the possible gauge groups, using the embedding
tensor, have been performed in $D=7$ \cite{Samtleben:2005bp} and $D=5$
\cite{deWit:2004nw}.


\section{Conclusions}\label{sec:conclusions}

In this paper we have given evidence that the de--forms that follow from $E_{11}$
and the  embedding tensors that follow from supersymmetric gauging are in
one-to-one correspondence to each other. In particular, we have shown that they
occur in precisely the same representations of the duality group. This is to be
expected assuming that the embedding tensor plays the role of the integration
constants of the de--form equations of motion. This requires a duality relation
of the following schematic form:

\begin{equation}
{}^\star\, G_{(D)} \sim \text{embedding tensor}\,,
\end{equation}
where $G_{(D)}$ is the curl of the de--form potential. The fact that the
representations coincide is a good hint that such a duality relation indeed
exists. It would be interesting to more precisely analyze this relation.

Finally, from the IIB case we know that not all top-forms couple to 1/2 BPS
branes. This is the reason that there is no quadruplet (but instead a non-linear
doublet) of IIB nine-branes given the fact that the D9-brane and its
$SL(2,\R)$ rotations couple to a quadruplet of 10--form potentials.
It would be interesting to see what the situation is for the top--forms in
$D<10$ dimensions. This would teach us something about the possible space--filling
branes in lower dimensions. The same applies to the de--forms and domain walls
in $D<10$ dimensions.


\acknowledgments

We would like to thank Joaquim Gomis and especially Diederik Roest for useful
discussions, and Mees de Roo and our other colleagues of the Groningen Journal
Club for discussions which triggered this project. We thank the Galileo Galilei
Institute for Theoretical Physics in Firenze for its hospitality and the INFN
for partial support.
E.B.~and T.N.~are supported by the European Commission FP6 program
MRTN-CT-2004-005104 in which E.B.~is associated to Utrecht University.
The work of E.B.~is partially supported by the Spanish grant BFM2003-01090
and by a Breedte Strategie grant of the University of Groningen.
The work of T.N.~ is part of the research programme of the ``Stichting voor
Fundamenteel Onderzoek der Materie (FOM)''.


\newpage

\appendix

\section{SimpLie: a simple program for Lie algebras}\label{app:simplie}

In this appendix we show how the root system of a Lie algebra $\g$ can be ordered
in representations of any of its subalgebras. We shall mainly be interested in
infinite-dimensional Kac-Moody algebras and their regular subalgebras. The
analysis follows the lines of \cite{Kleinschmidt:2003mf, Nicolai:2003fw, Damour:2002cu, Damour:2004zy}
but keep in mind it is only valid for algebras whose Cartan matrix is symmetric.
As we are dealing with infinite-dimensional algebras the resulting calculations
are rather cumbersome to work out by hand.

To automate the process we have written ``SimpLie'', which is a Java computer
program. The following sections explain the math behind it, and refer to specific
parts of the code where applicable. SimpLie and its source are available from
\cite{SimpLie}.

\subsection{Level decomposition}\label{app:ld}

Given a Lie algebra $\g$ with associated Cartan matrix $A$, we can form a regular
subalgebra $\s$ by deleting rows and columns from $A$. The resulting matrix is
then the Cartan matrix of the regular subalgebra of $\g$. This procedure
corresponds to deleting nodes from the associated Dynkin diagram of $\g$
\refSL{sl:dd}.

Consequently, a root $\alpha$ can be decomposed into contributions from the
deleted part and the regular subalgebra (repeated indices are summed over):

\begin{equation}
    \alpha = m^i \alpha_i = l^a \alpha_a + m^s \alpha_s .
\end{equation}
The values of the vector $l^a$ are more commonly called the levels of $\alpha$,
and $m^i$ the root labels. We have introduced the following indices:

\begin{align}
	i & = \{1, \ldots, r\},	& &\text{full algebra} \\
	a & = \{1, \ldots, n\},	& &\text{deleted nodes} \\
	s & = \{n+1, \ldots, r\},	& &\text{regular subalgebra}
\end{align}
where $r$ is the rank of $\g$ and $n$ is the number of deleted nodes.

Because the root system of $\g$ is graded with respect to $l^a$ and all the roots
of $\s$ have levels equal to zero, all roots of $\g$ with a particular $l^a$ form
highest weight modules of $\s$ under the adjoint action:

\begin{equation}
    [\s, \g_{l^a}] \subset \g_{l^a} .
\end{equation}
These modules are characterized by the Dynkin labels of their highest weight
state, which are given by

\begin{equation}
\label{A:dynkin_labels}
    p_s = A_{si} m^i = A_{st} m^t + A_{sa} l^a.
\end{equation}

For given levels $l^a$ it is easy to scan for possible valid highest weight
modules. They have to satisfy three conditions:

\begin{enumerate}[(i)]
	\item Their Dynkin labels all have to be integer and non-negative.
	\item \label{A:cond2} The $m^s$ have to be integers.
	\item \label{A:cond3} The length squared of the root must not exceed the maximum value.
\end{enumerate}
In the finite-dimensional case, the maximum value is given by the length squared
of the highest root. However, for Kac-Moody algebras there is no such thing as a
highest root. Fortunately, for the simply-laced cases the length squared of the
roots is bounded by $\alpha^2 \leq 2$ \cite{Kac:1990gs}. Thus we have

\begin{equation}
\label{A:root_length}
    \alpha^2
        = A_{ij} m^i m^j
        = S^{st} p_s p_t + \left( A_{ab} - S^{st} A_{sa} A_{tb} \right) l^a l^b \leq 2 ,
\end{equation}
where $S$ is the inverse of the Cartan matrix of $\s$. When $S$ has no negative
entries, the above formula is a monotonically increasing function of $p_s$ at
fixed levels. This is the case when $\s$ is finite.

To check condition \eqref{A:cond2} we can invert \eqref{A:dynkin_labels} to
obtain

\begin{equation}
\label{A:root_comp}
    m^s = S^{st} ( p_t - A_{ta} l^a) .
\end{equation}

When equations \eqref{A:root_comp} and \eqref{A:root_length} respectively satisfy
conditions \eqref{A:cond2} and \eqref{A:cond3} we have found a possible highest
weight representation of $\s$ \refSL{sl:ALS}.
Yet it remains to be seen whether or not this representation actually occurs
within $\g$. To that end, the root system of $\g$ must be constructed up to the
levels we are interested in. Furthermore, one has to check if a particular
representation occurs as a weight in another representation. These are the
subjects of the next sections.

Note that, instead of scanning for the highest weight, we can also look for the
lowest weight of the representation. It has Dynkin labels equal to minus the
Dynkin labels of the highest weight state of the conjugate representation. An
easy way to do this is to replace $p_s$ with $- p_s$ in the above analysis.
The advantage of this particular approach is that lowest weight states have
a lower height (given by $h = \sum_i m^i$), so that it suffices to construct
the root system of $\g$ to lower heights.

\subsection{Root system construction}\label{app:rs}

Assuming the root system of $\g$ has been constructed up to height $h$, there is
a simple procedure to determine all the roots of height $h+1$. Specifically one
considers, for all roots $\beta$ of height $h$, the string $s_{\alpha_i ; \beta}$
of simple roots $\alpha_i$ given by

\begin{equation}
	s_{\alpha_i ; \beta}
		= \{ \beta + k \alpha_i \ | \ k = - p, - p + 1, \ldots, q - 1, q \} ,
\end{equation}
where $p$ and $q$ satisfy

\begin{equation}
	\frac{ 2 ( \beta | \alpha_i ) }{ (\alpha_i | \alpha_i) }
		= p - q .
\end{equation}
In order for $\beta + \alpha_i$ to be a root, $q$ has to be positive. This is the
case when the inner product between $\beta$ and $\alpha_i$ is negative. If it is
positive, we have to perform a search through the previously generated roots in
the string to determine the value of $p$. Following this procedure, it is
possible to construct the root system to an arbitrary height starting from just
the simple roots \refSL{sl:rs}.

Knowing if a particular root $\alpha$ occurs in the root system of $\g$ is not
enough: one would also like to know its multiplicity $\textrm{mult}(\alpha)$.
The most straightforward method to calculate these multiplicities, albeit not
the most elegant one, is perhaps the Peterson recursion formula. It reads

\begin{equation}
\label{peterson}
	\Big( ( \alpha | \alpha ) - 2 h (\alpha) \Big) c_\alpha
		= \sum_{\substack{\alpha = \beta + \gamma \\ \beta,\gamma > 0}}
			(\beta | \gamma) c_\beta c_\gamma ,
\end{equation}
where the co-multiplicity $c_\alpha$ is given by

\begin{equation}
	c_\alpha
	= \sum_{k \geq 1} \frac 1 k \textrm{mult} \left( \frac \alpha k \right) .
\end{equation}
Factors for which $\alpha / k$ is not a root do not contribute to the sum.
The $\beta$ and $\gamma$ in \eqref{peterson} do not have to be roots but can also
be non-negative linear combinations of the simple roots. If they are not roots,
they have to be integer multiples of roots. Otherwise their co-multiplicity would
be zero and they would not contribute to the sum \eqref{peterson}.

\subsection{Outer multiplicities}

The number of times a representation of $\s$ actually occurs within the root
system of $\g$ is called its outer multiplicity $\mu$. In order to determine
$\mu$ we need to know the multiplicity of the representation as a weight in other
representations at the same level. Furthermore the multiplicity of the root
$\alpha$ in $\g$ associated to its highest weight state is needed. The outer
multiplicity $\mu$ then follows from \cite{Nicolai:2003fw}

\begin{equation}
\label{outer_mults}
	\textrm{mult}(\alpha) = \sum_i \mu(R_i) \textrm{mult}_{R_i} (\alpha),
\end{equation}
where $i$ runs over the number of representations at a fixed level, and $R_i$
is the $i$-th representation. The only unknowns remaining are the weight
multiplicities $\textrm{mult}_{R_i}$ of the representations of $\s$. These can be
calculated with the Freudenthal recursion formula, which reads

\begin{equation}
\label{freudenthal}
	\begin{split}
	\Bigl( (\Lambda | \Lambda) - 2 h (\Lambda)
		- (\lambda | \lambda) & - 2 h (\lambda) \Bigr) \textrm{mult}_{R(\Lambda)} (\lambda) \\
	& = 2 \sum_{\alpha > 0} \sum_{k \geq 1} ( \lambda + k\alpha | \alpha)
		\textrm{mult}_{R(\Lambda)} (\lambda + k\alpha) .
		\end{split}
\end{equation}
The first sum is over all positive roots of $\s$. The second terminates when
$\lambda + k \alpha$ leaves the weight system of $R(\Lambda)$, i.e. when the
height of the corresponding root exceeds that of $\Lambda$. In our case, the
highest weight $\Lambda$ is given by its Dynkin labels $p_s$ \refSL{sl:hwr}.

The procedure in SimpLie is the same at each level \refSL{sl:ALS}:
first all possible highest weight representations $R_i$ of $\s$ found by the
scanning procedure described in section \ref{app:ld} are gathered. Next we
calculate  $\textrm{mult}(\alpha)$ by the brute force method of section \ref{app:rs}.
Then all the relevant weights and their multiplicities of every $R_i$ are
calculated using, amongst others, the Freudenthal recursion formula. Finally,
the outer multiplicity of $R_i$ is determined using an iterative implementation
of \eqref{outer_mults}.

\section{Relevant Low Level Results} \label{app:results}

Here we list the output of SimpLie at low levels, using the various decompositions
of $E_{11}$ listed in table \ref{table1}. The regular subalgebra splits into a
part belonging to the gravity line $A_n$ (the white nodes) and a part belonging
to the internal duality group $G$ (the grey nodes).

In the following tables we respectively list the levels, the Dynkin labels of
$A_n$ and $G$, the root labels, the root length, the dimension of the
representations of $A_n$ and $G$, the multiplicity of the root, the outer
multiplicity, and the interpretation as a physical field. These physical fields
are also listed in table \ref{table2}. The de--forms and top--forms are indicated
by `de' and `top', respectively. When the internal group does not exist, we do
not list the corresponding columns. In all cases the Dynkin labels of the lowest
weights of the representations are given. The order of the Dynkin labels and the
root labels is determined by the numbering of the nodes in table \ref{table1}.
All tables are truncated at the point when the number of indices of the gravity
subalgebra representations exceed the dimension.

The interpretation of the representations at level zero as the graviton is,
unlike the $p$-forms at higher levels, not quite straightforward. The graviton
emerges when one combines the adjoint representation of $A_n$ with a scalar
coming from one of the deleted nodes, see \cite{Damour:2002cu,Kleinschmidt:2003mf}.
We have indicated these parts of the graviton by $\bar g_{\mu\nu}$ and
$\hat g_{\mu\nu}$, respectively.

\setlongtables
\footnotesize
\begin{longtable}{|r|r@{\ }r@{\ }r@{\ }r@{\ }r@{\ }r@{\ }r@{\ }r@{\ }r@{\ }r|r@{\ }r@{\ }r@{\ }r@{\ }r@{\ }r@{\ }r@{\ }r@{\ }r@{\ }r@{\ }r|r|r|r|r|c|}
\caption{$A_{10}$ representations in $E_{11}$ ($D=11$)} \\
\hline
\multicolumn{1}{|c|}{$l$} &
\multicolumn{10}{|c|}{$p_{\rm{grav}}$} &
\multicolumn{11}{|c|}{$m$} &
\multicolumn{1}{|c|}{$\alpha^2$} &
\multicolumn{1}{|c|}{$d_{\rm{reg}}$} &
\multicolumn{1}{|c|}{$\textrm{mult}(\alpha)$} &
\multicolumn{1}{|c|}{$\mu$} & fields \\
\hline
\hline
0 & 1 & 0 & 0 & 0 & 0 & 0 & 0 & 0 & 0 & 1 & 0 & -1 & -1 & -1 & -1 & -1 & -1 & -1 & -1 & -1 & -1 & 2 & 120 & 1 & 1 & $\bar g_{\mu\nu}$ \\
0 & 0 & 0 & 0 & 0 & 0 & 0 & 0 & 0 & 0 & 0 & 0 & 0 & 0 & 0 & 0 & 0 & 0 & 0 & 0 & 0 & 0 & 0 & 1 & 11 & 1 & $\hat g_{\mu\nu}$ \\
\hline
1 & 0 & 0 & 1 & 0 & 0 & 0 & 0 & 0 & 0 & 0 & 1 & 0 & 0 & 0 & 0 & 0 & 0 & 0 & 0 & 0 & 0 & 2 & 165 & 1 & 1 & $p=3$ \\
\hline
2 & 0 & 0 & 0 & 0 & 0 & 1 & 0 & 0 & 0 & 0 & 2 & 1 & 2 & 3 & 2 & 1 & 0 & 0 & 0 & 0 & 0 & 2 & 462 & 1 & 1 & ${}^\star\,  (p=3)$\\
\hline
3 & 1 & 0 & 0 & 0 & 0 & 0 & 0 & 1 & 0 & 0 & 3 & 1 & 3 & 5 & 4 & 3 & 2 & 1 & 0 & 0 & 0 & 2 & 1760 & 1 & 1 & ${}^\star\,  g_{\mu\nu}$\\
\hline
\end{longtable}

\begin{longtable}{|r@{\ }r|r@{\ }r@{\ }r@{\ }r@{\ }r@{\ }r@{\ }r@{\ }r@{\ }r|r@{\ }r@{\ }r@{\ }r@{\ }r@{\ }r@{\ }r@{\ }r@{\ }r@{\ }r@{\ }r|r|r|r|r|c|}
\caption{$A_9$ representations in $E_{11}$ (IIA)}  \\
\hline
\multicolumn{2}{|c|}{$l$} &
\multicolumn{9}{|c|}{$p_{\rm{grav}}$} &
\multicolumn{11}{|c|}{$m$} &
\multicolumn{1}{|c|}{$\alpha^2$} &
\multicolumn{1}{|c|}{$d_{\rm{reg}}$} &
\multicolumn{1}{|c|}{$\textrm{mult}(\alpha)$} &
\multicolumn{1}{|c|}{$\mu$} & fields \\
\hline
\hline
0 & 0 & 1 & 0 & 0 & 0 & 0 & 0 & 0 & 0 & 1 & 0 & 0 & -1 & -1 & -1 & -1 & -1 & -1 & -1 & -1 & -1 & 2 & 99 & 1 & 1 & $\bar g_{\mu\nu}$ \\
0 & 0 & 0 & 0 & 0 & 0 & 0 & 0 & 0 & 0 & 0 & 0 & 0 & 0 & 0 & 0 & 0 & 0 & 0 & 0 & 0 & 0 & 0 & 1 & 11 & 2 & $p=0, \hat g_{\mu\nu}$\\
\hline
1 & 0 & 0 & 1 & 0 & 0 & 0 & 0 & 0 & 0 & 0 & 1 & 0 & 0 & 0 & 0 & 0 & 0 & 0 & 0 & 0 & 0 & 2 & 45 & 1 & 1 & $p=2$ \\
\hline
0 & 1 & 1 & 0 & 0 & 0 & 0 & 0 & 0 & 0 & 0 & 0 & 1 & 0 & 0 & 0 & 0 & 0 & 0 & 0 & 0 & 0 & 2 & 10 & 1 & 1 & $p=1$ \\
\hline
1 & 1 & 0 & 0 & 1 & 0 & 0 & 0 & 0 & 0 & 0 & 1 & 1 & 1 & 1 & 0 & 0 & 0 & 0 & 0 & 0 & 0 & 2 & 120 & 1 & 1 & $p=3$ \\
\hline
2 & 1 & 0 & 0 & 0 & 0 & 1 & 0 & 0 & 0 & 0 & 2 & 1 & 2 & 3 & 2 & 1 & 0 & 0 & 0 & 0 & 0 & 2 & 252 & 1 & 1 & ${}^\star\,  (p=3)$ \\
\hline
3 & 1 & 0 & 0 & 0 & 0 & 0 & 0 & 1 & 0 & 0 & 3 & 1 & 3 & 5 & 4 & 3 & 2 & 1 & 0 & 0 & 0 & 2 & 120 & 1 & 1 & ${}^\star\,  (p=1)$ \\
\hline
4 & 1 & 0 & 0 & 0 & 0 & 0 & 0 & 0 & 0 & 1 & 4 & 1 & 4 & 7 & 6 & 5 & 4 & 3 & 2 & 1 & 0 & 2 & 10 & 1 & 1 & $\textrm{de}$ \\
\hline
2 & 2 & 0 & 0 & 0 & 0 & 0 & 1 & 0 & 0 & 0 & 2 & 2 & 3 & 4 & 3 & 2 & 1 & 0 & 0 & 0 & 0 & 2 & 210 & 1 & 1 & ${}^\star\,  (p=2)$ \\
\hline
3 & 2 & 1 & 0 & 0 & 0 & 0 & 0 & 1 & 0 & 0 & 3 & 2 & 3 & 5 & 4 & 3 & 2 & 1 & 0 & 0 & 0 & 2 & 1155 & 1 & 1 & ${}^\star\,  g_{\mu\nu}$ \\
3 & 2 & 0 & 0 & 0 & 0 & 0 & 0 & 0 & 1 & 0 & 3 & 2 & 4 & 6 & 5 & 4 & 3 & 2 & 1 & 0 & 0 & 0 & 45 & 8 & 1 & ${}^\star\,  (p=0)$ \\
\hline
4 & 2 & 0 & 1 & 0 & 0 & 0 & 0 & 0 & 1 & 0 & 4 & 2 & 4 & 6 & 5 & 4 & 3 & 2 & 1 & 0 & 0 & 2 & 1925 & 1 & 1 & \\
4 & 2 & 1 & 0 & 0 & 0 & 0 & 0 & 0 & 0 & 1 & 4 & 2 & 4 & 7 & 6 & 5 & 4 & 3 & 2 & 1 & 0 & 0 & 99 & 8 & 1 & \\
4 & 2 & 0 & 0 & 0 & 0 & 0 & 0 & 0 & 0 & 0 & 4 & 2 & 5 & 8 & 7 & 6 & 5 & 4 & 3 & 2 & 1 & -2 & 1 & 46 & 2 & $\textrm{top}$ \\
\hline
3 & 3 & 1 & 0 & 0 & 0 & 0 & 0 & 0 & 1 & 0 & 3 & 3 & 4 & 6 & 5 & 4 & 3 & 2 & 1 & 0 & 0 & 2 & 440 & 1 & 1 & \\
\hline
\end{longtable}

\begin{longtable}{|r|r@{\ }r@{\ }r@{\ }r@{\ }r@{\ }r@{\ }r@{\ }r@{\ }r|r|r@{\ }r@{\ }r@{\ }r@{\ }r@{\ }r@{\ }r@{\ }r@{\ }r@{\ }r@{\ }r|r|r|r|r|r|c|}
\caption{$A_9 \times A_1$ representations in $E_{11}$ (IIB)} \\
\hline
\multicolumn{1}{|c|}{$l$} &
\multicolumn{9}{|c|}{$p_{\rm{grav}}$} &
\multicolumn{1}{|c|}{$p_{\rm{int}}$} &
\multicolumn{11}{|c|}{$m$} &
\multicolumn{1}{|c|}{$\alpha^2$} &
\multicolumn{1}{|c|}{$d_{\rm{reg}}$} &
\multicolumn{1}{|c|}{$d_{\rm{int}}$} &
\multicolumn{1}{|c|}{$\textrm{mult}(\alpha)$} &
\multicolumn{1}{|c|}{$\mu$} & fields \\
\hline
\hline
0 & 1 & 0 & 0 & 0 & 0 & 0 & 0 & 0 & 1 & 0 & -1 & 0 & 0 & -1 & -1 & -1 & -1 & -1 & -1 & -1 & -1 & 2 & 99 & 1 & 1 & 1 & $\bar g_{\mu\nu}$ \\
0 & 0 & 0 & 0 & 0 & 0 & 0 & 0 & 0 & 0 & 2 & 0 & -1 & 0 & 0 & 0 & 0 & 0 & 0 & 0 & 0 & 0 & 2 & 1 & 3 & 1 & 1 & $p=0$ \\
0 & 0 & 0 & 0 & 0 & 0 & 0 & 0 & 0 & 0 & 0 & 0 & 0 & 0 & 0 & 0 & 0 & 0 & 0 & 0 & 0 & 0 & 0 & 1 & 1 & 11 & 1 & $\hat g_{\mu\nu}$ \\
\hline
1 & 0 & 1 & 0 & 0 & 0 & 0 & 0 & 0 & 0 & 1 & 0 & 0 & 1 & 0 & 0 & 0 & 0 & 0 & 0 & 0 & 0 & 2 & 45 & 2 & 1 & 1 & $p=2$ \\
\hline
2 & 0 & 0 & 0 & 1 & 0 & 0 & 0 & 0 & 0 & 0 & 1 & 1 & 2 & 2 & 1 & 0 & 0 & 0 & 0 & 0 & 0 & 2 & 210 & 1 & 1 & 1 & $p=4$\\
\hline
3 & 0 & 0 & 0 & 0 & 0 & 1 & 0 & 0 & 0 & 1 & 2 & 1 & 3 & 4 & 3 & 2 & 1 & 0 & 0 & 0 & 0 & 2 & 210 & 2 & 1 & 1 & ${}^\star\,  (p=2)$\\
\hline
4 & 1 & 0 & 0 & 0 & 0 & 0 & 1 & 0 & 0 & 0 & 2 & 2 & 4 & 5 & 4 & 3 & 2 & 1 & 0 & 0 & 0 & 2 & 1155 & 1 & 1 & 1 & ${}^\star\,  g_{\mu\nu}$ \\
4 & 0 & 0 & 0 & 0 & 0 & 0 & 0 & 1 & 0 & 2 & 3 & 1 & 4 & 6 & 5 & 4 & 3 & 2 & 1 & 0 & 0 & 2 & 45 & 3 & 1 & 1 & ${}^\star\,  (p=0)$ \\
\hline
5 & 0 & 1 & 0 & 0 & 0 & 0 & 0 & 1 & 0 & 1 & 3 & 2 & 5 & 6 & 5 & 4 & 3 & 2 & 1 & 0 & 0 & 2 & 1925 & 2 & 1 & 1 & \\
5 & 1 & 0 & 0 & 0 & 0 & 0 & 0 & 0 & 1 & 1 & 3 & 2 & 5 & 7 & 6 & 5 & 4 & 3 & 2 & 1 & 0 & 0 & 99 & 2 & 8 & 1 & \\
5 & 0 & 0 & 0 & 0 & 0 & 0 & 0 & 0 & 0 & 3 & 4 & 1 & 5 & 8 & 7 & 6 & 5 & 4 & 3 & 2 & 1 & 2 & 1 & 4 & 1 & 1 & $\textrm{top}$ \\
5 & 0 & 0 & 0 & 0 & 0 & 0 & 0 & 0 & 0 & 1 & 4 & 2 & 5 & 8 & 7 & 6 & 5 & 4 & 3 & 2 & 1 & -2 & 1 & 2 & 46 & 1 & $\textrm{top}$ \\
\hline
\end{longtable}

\begin{longtable}{|r@{\ }r|r@{\ }r@{\ }r@{\ }r@{\ }r@{\ }r@{\ }r@{\ }r|r|r@{\ }r@{\ }r@{\ }r@{\ }r@{\ }r@{\ }r@{\ }r@{\ }r@{\ }r@{\ }r|r|r|r|r|r|c|}
\caption{$A_8 \times A_1$ representations in $E_{11}$ ($D=9$)} \\
\hline
\multicolumn{2}{|c|}{$l$} &
\multicolumn{8}{|c|}{$p_{\rm{grav}}$} &
\multicolumn{1}{|c|}{$p_{\rm{int}}$} &
\multicolumn{11}{|c|}{$m$} &
\multicolumn{1}{|c|}{$\alpha^2$} &
\multicolumn{1}{|c|}{$d_{\rm{reg}}$} &
\multicolumn{1}{|c|}{$d_{\rm{int}}$} &
\multicolumn{1}{|c|}{$\textrm{mult}(\alpha)$} &
\multicolumn{1}{|c|}{$\mu$} & fields \\
\hline
\hline
0 & 0 & 1 & 0 & 0 & 0 & 0 & 0 & 0 & 1 & 0 & 0 & 0 & 0 & -1 & -1 & -1 & -1 & -1 & -1 & -1 & -1 & 2 & 80 & 1 & 1 & 1 & $\bar g_{\mu\nu}$ \\
0 & 0 & 0 & 0 & 0 & 0 & 0 & 0 & 0 & 0 & 2 & 0 & -1 & 0 & 0 & 0 & 0 & 0 & 0 & 0 & 0 & 0 & 2 & 1 & 3 & 1 & 1 & $p=0$ \\
0 & 0 & 0 & 0 & 0 & 0 & 0 & 0 & 0 & 0 & 0 & 0 & 0 & 0 & 0 & 0 & 0 & 0 & 0 & 0 & 0 & 0 & 0 & 1 & 1 & 11 & 2 & $p=0, \hat g_{\mu\nu}$\\
\hline
1 & 0 & 1 & 0 & 0 & 0 & 0 & 0 & 0 & 0 & 0 & 1 & 0 & 0 & 0 & 0 & 0 & 0 & 0 & 0 & 0 & 0 & 2 & 9 & 1 & 1 & 1 & $p=1$ \\
\hline
0 & 1 & 1 & 0 & 0 & 0 & 0 & 0 & 0 & 0 & 1 & 0 & 0 & 1 & 0 & 0 & 0 & 0 & 0 & 0 & 0 & 0 & 2 & 9 & 2 & 1 & 1 & $p=1$ \\
\hline
1 & 1 & 0 & 1 & 0 & 0 & 0 & 0 & 0 & 0 & 1 & 1 & 0 & 1 & 1 & 0 & 0 & 0 & 0 & 0 & 0 & 0 & 2 & 36 & 2 & 1 & 1 & $p=2$ \\
\hline
1 & 2 & 0 & 0 & 1 & 0 & 0 & 0 & 0 & 0 & 0 & 1 & 1 & 2 & 2 & 1 & 0 & 0 & 0 & 0 & 0 & 0 & 2 & 84 & 1 & 1 & 1 & $p=3$ \\
\hline
2 & 2 & 0 & 0 & 0 & 1 & 0 & 0 & 0 & 0 & 0 & 2 & 1 & 2 & 3 & 2 & 1 & 0 & 0 & 0 & 0 & 0 & 2 & 126 & 1 & 1 & 1 & ${}^\star\,  (p=3)$ \\
\hline
2 & 3 & 0 & 0 & 0 & 0 & 1 & 0 & 0 & 0 & 1 & 2 & 1 & 3 & 4 & 3 & 2 & 1 & 0 & 0 & 0 & 0 & 2 & 126 & 2 & 1 & 1 & ${}^\star\,  (p=2)$ \\
\hline
3 & 3 & 0 & 0 & 0 & 0 & 0 & 1 & 0 & 0 & 1 & 3 & 1 & 3 & 5 & 4 & 3 & 2 & 1 & 0 & 0 & 0 & 2 & 84 & 2 & 1 & 1 & ${}^\star\,  (p=1)$ \\
\hline
2 & 4 & 0 & 0 & 0 & 0 & 0 & 1 & 0 & 0 & 0 & 2 & 2 & 4 & 5 & 4 & 3 & 2 & 1 & 0 & 0 & 0 & 2 & 84 & 1 & 1 & 1 & ${}^\star\,  (p=1)$ \\
\hline
3 & 4 & 1 & 0 & 0 & 0 & 0 & 1 & 0 & 0 & 0 & 3 & 2 & 4 & 5 & 4 & 3 & 2 & 1 & 0 & 0 & 0 & 2 & 720 & 1 & 1 & 1 & ${}^\star\,  g_{\mu\nu}$ \\
3 & 4 & 0 & 0 & 0 & 0 & 0 & 0 & 1 & 0 & 2 & 3 & 1 & 4 & 6 & 5 & 4 & 3 & 2 & 1 & 0 & 0 & 2 & 36 & 3 & 1 & 1 & ${}^\star\,  (p=0)$ \\
3 & 4 & 0 & 0 & 0 & 0 & 0 & 0 & 1 & 0 & 0 & 3 & 2 & 4 & 6 & 5 & 4 & 3 & 2 & 1 & 0 & 0 & 0 & 36 & 1 & 8 & 1 & ${}^\star\,  (p=0)$ \\
\hline
4 & 4 & 1 & 0 & 0 & 0 & 0 & 0 & 1 & 0 & 0 & 4 & 2 & 4 & 6 & 5 & 4 & 3 & 2 & 1 & 0 & 0 & 2 & 315 & 1 & 1 & 1 & \\
4 & 4 & 0 & 0 & 0 & 0 & 0 & 0 & 0 & 1 & 2 & 4 & 1 & 4 & 7 & 6 & 5 & 4 & 3 & 2 & 1 & 0 & 2 & 9 & 3 & 1 & 1 & $\textrm{de}$ \\
\hline
3 & 5 & 1 & 0 & 0 & 0 & 0 & 0 & 1 & 0 & 1 & 3 & 2 & 5 & 6 & 5 & 4 & 3 & 2 & 1 & 0 & 0 & 2 & 315 & 2 & 1 & 1 & \\
3 & 5 & 0 & 0 & 0 & 0 & 0 & 0 & 0 & 1 & 1 & 3 & 2 & 5 & 7 & 6 & 5 & 4 & 3 & 2 & 1 & 0 & 0 & 9 & 2 & 8 & 1 & $\textrm{de}$ \\
\hline
4 & 5 & 0 & 1 & 0 & 0 & 0 & 0 & 1 & 0 & 1 & 4 & 2 & 5 & 7 & 5 & 4 & 3 & 2 & 1 & 0 & 0 & 2 & 1215 & 2 & 1 & 1 & \\
4 & 5 & 1 & 0 & 0 & 0 & 0 & 0 & 0 & 1 & 1 & 4 & 2 & 5 & 7 & 6 & 5 & 4 & 3 & 2 & 1 & 0 & 0 & 80 & 2 & 8 & 2 & \\
4 & 5 & 0 & 0 & 0 & 0 & 0 & 0 & 0 & 0 & 3 & 4 & 1 & 5 & 8 & 7 & 6 & 5 & 4 & 3 & 2 & 1 & 2 & 1 & 4 & 1 & 1 & $\textrm{top}$ \\
4 & 5 & 0 & 0 & 0 & 0 & 0 & 0 & 0 & 0 & 1 & 4 & 2 & 5 & 8 & 7 & 6 & 5 & 4 & 3 & 2 & 1 & -2 & 1 & 2 & 46 & 2 & $\textrm{top}$ \\
\hline
3 & 6 & 1 & 0 & 0 & 0 & 0 & 0 & 0 & 1 & 0 & 3 & 3 & 6 & 7 & 6 & 5 & 4 & 3 & 2 & 1 & 0 & 2 & 80 & 1 & 1 & 1 & \\
\hline
\end{longtable}

\begin{longtable}{|r|r@{\ }r@{\ }r@{\ }r@{\ }r@{\ }r@{\ }r|r@{\ }r@{\ }r|r@{\ }r@{\ }r@{\ }r@{\ }r@{\ }r@{\ }r@{\ }r@{\ }r@{\ }r@{\ }r|r|r|r|r|r|c|}
\caption{$A_7 \times ( A_2 \times A_1 )$ representations in $E_{11}$ ($D=8$)} \\
\hline
\multicolumn{1}{|c|}{$l$} &
\multicolumn{7}{|c|}{$p_{\rm{grav}}$} &
\multicolumn{3}{|c|}{$p_{\rm{int}}$} &
\multicolumn{11}{|c|}{$m$} &
\multicolumn{1}{|c|}{$\alpha^2$} &
\multicolumn{1}{|c|}{$d_{\rm{reg}}$} &
\multicolumn{1}{|c|}{$d_{\rm{int}}$} &
\multicolumn{1}{|c|}{$\textrm{mult}(\alpha)$} &
\multicolumn{1}{|c|}{$\mu$} & fields \\
\hline
\hline
0 & 1 & 0 & 0 & 0 & 0 & 0 & 1 & 0 & 0 & 0 & 0 & 0 & 0 & 0 & -1 & -1 & -1 & -1 & -1 & -1 & -1 & 2 & 63 & 1 & 1 & 1 & $\bar g_{\mu\nu}$ \\
0 & 0 & 0 & 0 & 0 & 0 & 0 & 0 & 0 & 1 & 1 & 0 & -1 & -1 & 0 & 0 & 0 & 0 & 0 & 0 & 0 & 0 & 2 & 1 & 8 & 1 & 1 & $p=0$ \\
0 & 0 & 0 & 0 & 0 & 0 & 0 & 0 & 2 & 0 & 0 & -1 & 0 & 0 & 0 & 0 & 0 & 0 & 0 & 0 & 0 & 0 & 2 & 1 & 3 & 1 & 1 & $p=0$ \\
0 & 0 & 0 & 0 & 0 & 0 & 0 & 0 & 0 & 0 & 0 & 0 & 0 & 0 & 0 & 0 & 0 & 0 & 0 & 0 & 0 & 0 & 0 & 1 & 1 & 11 & 1 & $\hat g_{\mu\nu}$ \\
\hline
1 & 1 & 0 & 0 & 0 & 0 & 0 & 0 & 1 & 0 & 1 & 0 & 0 & 0 & 1 & 0 & 0 & 0 & 0 & 0 & 0 & 0 & 2 & 8 & 6 & 1 & 1 & $p=1$ \\
\hline
2 & 0 & 1 & 0 & 0 & 0 & 0 & 0 & 0 & 1 & 0 & 1 & 0 & 1 & 2 & 1 & 0 & 0 & 0 & 0 & 0 & 0 & 2 & 28 & 3 & 1 & 1 & $p=2$ \\
\hline
3 & 0 & 0 & 1 & 0 & 0 & 0 & 0 & 1 & 0 & 0 & 1 & 1 & 2 & 3 & 2 & 1 & 0 & 0 & 0 & 0 & 0 & 2 & 56 & 2 & 1 & 1 & $p=3,$ \\
 &  &  &  &  &  &  &  &  &  &  &  &  &  &  &  &  &  &  &  &  &  &  &  & &  &  & ${}^\star\, (p=3)$ \\
\hline
4 & 0 & 0 & 0 & 1 & 0 & 0 & 0 & 0 & 0 & 1 & 2 & 1 & 2 & 4 & 3 & 2 & 1 & 0 & 0 & 0 & 0 & 2 & 70 & 3 & 1 & 1 & ${}^\star\,  (p=2)$ \\
\hline
5 & 0 & 0 & 0 & 0 & 1 & 0 & 0 & 1 & 1 & 0 & 2 & 1 & 3 & 5 & 4 & 3 & 2 & 1 & 0 & 0 & 0 & 2 & 56 & 6 & 1 & 1 & ${}^\star\,  (p=1)$ \\
\hline
6 & 1 & 0 & 0 & 0 & 1 & 0 & 0 & 0 & 0 & 0 & 3 & 2 & 4 & 6 & 4 & 3 & 2 & 1 & 0 & 0 & 0 & 2 & 420 & 1 & 1 & 1 & ${}^\star\,  g_{\mu\nu}$ \\
6 & 0 & 0 & 0 & 0 & 0 & 1 & 0 & 0 & 1 & 1 & 3 & 1 & 3 & 6 & 5 & 4 & 3 & 2 & 1 & 0 & 0 & 2 & 28 & 8 & 1 & 1 & ${}^\star\,  (p=0)$ \\
6 & 0 & 0 & 0 & 0 & 0 & 1 & 0 & 2 & 0 & 0 & 2 & 2 & 4 & 6 & 5 & 4 & 3 & 2 & 1 & 0 & 0 & 2 & 28 & 3 & 1 & 1 & ${}^\star\,  (p=0)$ \\
\hline
7 & 1 & 0 & 0 & 0 & 0 & 1 & 0 & 1 & 0 & 1 & 3 & 2 & 4 & 7 & 5 & 4 & 3 & 2 & 1 & 0 & 0 & 2 & 216 & 6 & 1 & 1 & \\
7 & 0 & 0 & 0 & 0 & 0 & 0 & 1 & 1 & 2 & 0 & 3 & 1 & 4 & 7 & 6 & 5 & 4 & 3 & 2 & 1 & 0 & 2 & 8 & 12 & 1 & 1 & $\textrm{de}$ \\
7 & 0 & 0 & 0 & 0 & 0 & 0 & 1 & 1 & 0 & 1 & 3 & 2 & 4 & 7 & 6 & 5 & 4 & 3 & 2 & 1 & 0 & 0 & 8 & 6 & 8 & 1 & $\textrm{de}$ \\
\hline
8 & 0 & 1 & 0 & 0 & 0 & 1 & 0 & 0 & 1 & 0 & 4 & 2 & 5 & 8 & 6 & 4 & 3 & 2 & 1 & 0 & 0 & 2 & 720 & 3 & 1 & 1 & \\
8 & 1 & 0 & 0 & 0 & 0 & 0 & 1 & 0 & 0 & 2 & 4 & 2 & 4 & 8 & 6 & 5 & 4 & 3 & 2 & 1 & 0 & 2 & 63 & 6 & 1 & 1 & \\
8 & 1 & 0 & 0 & 0 & 0 & 0 & 1 & 2 & 1 & 0 & 3 & 2 & 5 & 8 & 6 & 5 & 4 & 3 & 2 & 1 & 0 & 2 & 63 & 9 & 1 & 1 & \\
8 & 1 & 0 & 0 & 0 & 0 & 0 & 1 & 0 & 1 & 0 & 4 & 2 & 5 & 8 & 6 & 5 & 4 & 3 & 2 & 1 & 0 & 0 & 63 & 3 & 8 & 1 & \\
8 & 0 & 0 & 0 & 0 & 0 & 0 & 0 & 0 & 2 & 1 & 4 & 1 & 4 & 8 & 7 & 6 & 5 & 4 & 3 & 2 & 1 & 2 & 1 & 15 & 1 & 1 & $\textrm{top}$ \\
8 & 0 & 0 & 0 & 0 & 0 & 0 & 0 & 2 & 1 & 0 & 3 & 2 & 5 & 8 & 7 & 6 & 5 & 4 & 3 & 2 & 1 & 0 & 1 & 9 & 8 & 1 & $\textrm{top}$ \\
8 & 0 & 0 & 0 & 0 & 0 & 0 & 0 & 0 & 1 & 0 & 4 & 2 & 5 & 8 & 7 & 6 & 5 & 4 & 3 & 2 & 1 & -2 & 1 & 3 & 46 & 2 & $\textrm{top}$ \\
\hline
\end{longtable}

\begin{longtable}{|r|r@{\ }r@{\ }r@{\ }r@{\ }r@{\ }r|r@{\ }r@{\ }r@{\ }r|r@{\ }r@{\ }r@{\ }r@{\ }r@{\ }r@{\ }r@{\ }r@{\ }r@{\ }r@{\ }r|r|r|r|r|r|c|}
\caption{$A_6 \times A_4$ representations in $E_{11}$ ($D=7$)} \\
\hline
\multicolumn{1}{|c|}{$l$} &
\multicolumn{6}{|c|}{$p_{\rm{grav}}$} &
\multicolumn{4}{|c|}{$p_{\rm{int}}$} &
\multicolumn{11}{|c|}{$m$} &
\multicolumn{1}{|c|}{$\alpha^2$} &
\multicolumn{1}{|c|}{$d_{\rm{reg}}$} &
\multicolumn{1}{|c|}{$d_{\rm{int}}$} &
\multicolumn{1}{|c|}{$\textrm{mult}(\alpha)$} &
\multicolumn{1}{|c|}{$\mu$} & fields \\
\hline
\hline
0 & 1 & 0 & 0 & 0 & 0 & 1 & 0 & 0 & 0 & 0 & 0 & 0 & 0 & 0 & 0 & -1 & -1 & -1 & -1 & -1 & -1 & 2 & 48 & 1 & 1 & 1 & $\bar g_{\mu\nu}$ \\
0 & 0 & 0 & 0 & 0 & 0 & 0 & 1 & 1 & 0 & 0 & -1 & -1 & -1 & -1 & 0 & 0 & 0 & 0 & 0 & 0 & 0 & 2 & 1 & 24 & 1 & 1 & $p=0$ \\
0 & 0 & 0 & 0 & 0 & 0 & 0 & 0 & 0 & 0 & 0 & 0 & 0 & 0 & 0 & 0 & 0 & 0 & 0 & 0 & 0 & 0 & 0 & 1 & 1 & 11 & 1 & $\hat g_{\mu\nu}$ \\
\hline
1 & 1 & 0 & 0 & 0 & 0 & 0 & 0 & 0 & 0 & 1 & 0 & 0 & 0 & 0 & 1 & 0 & 0 & 0 & 0 & 0 & 0 & 2 & 7 & 10 & 1 & 1 & $p=1$ \\
\hline
2 & 0 & 1 & 0 & 0 & 0 & 0 & 0 & 1 & 0 & 0 & 1 & 0 & 1 & 2 & 2 & 1 & 0 & 0 & 0 & 0 & 0 & 2 & 21 & 5 & 1 & 1 & $p=2$ \\
\hline
3 & 0 & 0 & 1 & 0 & 0 & 0 & 1 & 0 & 0 & 0 & 1 & 1 & 2 & 3 & 3 & 2 & 1 & 0 & 0 & 0 & 0 & 2 & 35 & 5 & 1 & 1 & ${}^\star\,  (p=2)$ \\
\hline
4 & 0 & 0 & 0 & 1 & 0 & 0 & 0 & 0 & 1 & 0 & 2 & 1 & 2 & 4 & 4 & 3 & 2 & 1 & 0 & 0 & 0 & 2 & 35 & 10 & 1 & 1 & ${}^\star\,  (p=1)$ \\
\hline
5 & 1 & 0 & 0 & 1 & 0 & 0 & 0 & 0 & 0 & 0 & 3 & 2 & 4 & 6 & 5 & 3 & 2 & 1 & 0 & 0 & 0 & 2 & 224 & 1 & 1 & 1 & ${}^\star\,  g_{\mu\nu}$ \\
5 & 0 & 0 & 0 & 0 & 1 & 0 & 1 & 1 & 0 & 0 & 2 & 1 & 3 & 5 & 5 & 4 & 3 & 2 & 1 & 0 & 0 & 2 & 21 & 24 & 1 & 1 & ${}^\star\,  (p=0)$ \\
\hline
6 & 1 & 0 & 0 & 0 & 1 & 0 & 0 & 0 & 0 & 1 & 3 & 2 & 4 & 6 & 6 & 4 & 3 & 2 & 1 & 0 & 0 & 2 & 140 & 10 & 1 & 1 & \\
6 & 0 & 0 & 0 & 0 & 0 & 1 & 0 & 1 & 1 & 0 & 3 & 1 & 3 & 6 & 6 & 5 & 4 & 3 & 2 & 1 & 0 & 2 & 7 & 40 & 1 & 1 & $\textrm{de}$ \\
6 & 0 & 0 & 0 & 0 & 0 & 1 & 2 & 0 & 0 & 0 & 2 & 2 & 4 & 6 & 6 & 5 & 4 & 3 & 2 & 1 & 0 & 2 & 7 & 15 & 1 & 1 & $\textrm{de}$ \\
\hline
7 & 0 & 1 & 0 & 0 & 1 & 0 & 0 & 1 & 0 & 0 & 4 & 2 & 5 & 8 & 7 & 5 & 3 & 2 & 1 & 0 & 0 & 2 & 392 & 5 & 1 & 1 & \\
7 & 1 & 0 & 0 & 0 & 0 & 1 & 1 & 0 & 1 & 0 & 3 & 2 & 4 & 7 & 7 & 5 & 4 & 3 & 2 & 1 & 0 & 2 & 48 & 45 & 1 & 1 & \\
7 & 1 & 0 & 0 & 0 & 0 & 1 & 0 & 1 & 0 & 0 & 4 & 2 & 5 & 8 & 7 & 5 & 4 & 3 & 2 & 1 & 0 & 0 & 48 & 5 & 8 & 1 & \\
7 & 0 & 0 & 0 & 0 & 0 & 0 & 1 & 2 & 0 & 0 & 3 & 1 & 4 & 7 & 7 & 6 & 5 & 4 & 3 & 2 & 1 & 2 & 1 & 70 & 1 & 1 & $\textrm{top}$ \\
7 & 0 & 0 & 0 & 0 & 0 & 0 & 1 & 0 & 1 & 0 & 3 & 2 & 4 & 7 & 7 & 6 & 5 & 4 & 3 & 2 & 1 & 0 & 1 & 45 & 8 & 1 & $\textrm{top}$ \\
7 & 0 & 0 & 0 & 0 & 0 & 0 & 0 & 1 & 0 & 0 & 4 & 2 & 5 & 8 & 7 & 6 & 5 & 4 & 3 & 2 & 1 & -2 & 1 & 5 & 46 & 1 & $\textrm{top}$ \\
\hline
\end{longtable}

\begin{longtable}{|r|r@{\ }r@{\ }r@{\ }r@{\ }r|r@{\ }r@{\ }r@{\ }r@{\ }r|r@{\ }r@{\ }r@{\ }r@{\ }r@{\ }r@{\ }r@{\ }r@{\ }r@{\ }r@{\ }r|r|r|r|r|r|c|}
\caption{$A_5 \times E_5$ representations in $E_{11}$ ($D=6$)} \\
\hline
\multicolumn{1}{|c|}{$l$} &
\multicolumn{5}{|c|}{$p_{\rm{grav}}$} &
\multicolumn{5}{|c|}{$p_{\rm{int}}$} &
\multicolumn{11}{|c|}{$m$} &
\multicolumn{1}{|c|}{$\alpha^2$} &
\multicolumn{1}{|c|}{$d_{\rm{reg}}$} &
\multicolumn{1}{|c|}{$d_{\rm{int}}$} &
\multicolumn{1}{|c|}{$\textrm{mult}(\alpha)$} &
\multicolumn{1}{|c|}{$\mu$} & fields \\
\hline
\hline
0 & 0 & 0 & 0 & 0 & 0 & 0 & 0 & 1 & 0 & 0 & -1 & -1 & -2 & -2 & -1 & 0 & 0 & 0 & 0 & 0 & 0 & 2 & 1 & 45 & 1 & 1 & $p=0$ \\
0 & 1 & 0 & 0 & 0 & 1 & 0 & 0 & 0 & 0 & 0 & 0 & 0 & 0 & 0 & 0 & 0 & -1 & -1 & -1 & -1 & -1 & 2 & 35 & 1 & 1 & 1 & $\bar g_{\mu\nu}$ \\
0 & 0 & 0 & 0 & 0 & 0 & 0 & 0 & 0 & 0 & 0 & 0 & 0 & 0 & 0 & 0 & 0 & 0 & 0 & 0 & 0 & 0 & 0 & 1 & 1 & 11 & 1 & $\hat g_{\mu\nu}$ \\
\hline
1 & 1 & 0 & 0 & 0 & 0 & 0 & 0 & 0 & 0 & 1 & 0 & 0 & 0 & 0 & 0 & 1 & 0 & 0 & 0 & 0 & 0 & 2 & 6 & 16 & 1 & 1 & $p=1$ \\
\hline
2 & 0 & 1 & 0 & 0 & 0 & 0 & 1 & 0 & 0 & 0 & 1 & 0 & 1 & 2 & 2 & 2 & 1 & 0 & 0 & 0 & 0 & 2 & 15 & 10 & 1 & 1 & $p=2,$ \\
 &  &  &  &  &  &  &  &  &  &  &  &  &  &  &  &  &  &  &  &  &  &  &  & &  &  & ${}^\star\, (p=2)$ \\
\hline
3 & 0 & 0 & 1 & 0 & 0 & 1 & 0 & 0 & 0 & 0 & 1 & 1 & 2 & 3 & 3 & 3 & 2 & 1 & 0 & 0 & 0 & 2 & 20 & 16 & 1 & 1 & ${}^\star\,  (p=1)$ \\
\hline
4 & 0 & 0 & 0 & 1 & 0 & 0 & 0 & 1 & 0 & 0 & 2 & 1 & 2 & 4 & 4 & 4 & 3 & 2 & 1 & 0 & 0 & 2 & 15 & 45 & 1 & 1 & ${}^\star\,  (p=0)$ \\
4 & 1 & 0 & 1 & 0 & 0 & 0 & 0 & 0 & 0 & 0 & 3 & 2 & 4 & 6 & 5 & 4 & 2 & 1 & 0 & 0 & 0 & 2 & 105 & 1 & 1 & 1 & ${}^\star\,  g_{\mu\nu}$ \\
\hline
5 & 1 & 0 & 0 & 1 & 0 & 0 & 0 & 0 & 0 & 1 & 3 & 2 & 4 & 6 & 5 & 5 & 3 & 2 & 1 & 0 & 0 & 2 & 84 & 16 & 1 & 1 & \\
5 & 0 & 0 & 0 & 0 & 1 & 1 & 1 & 0 & 0 & 0 & 2 & 1 & 3 & 5 & 5 & 5 & 4 & 3 & 2 & 1 & 0 & 2 & 6 & 144 & 1 & 1 & $\textrm{de}$ \\
\hline
6 & 1 & 0 & 0 & 0 & 1 & 0 & 0 & 0 & 1 & 0 & 3 & 2 & 4 & 6 & 6 & 6 & 4 & 3 & 2 & 1 & 0 & 2 & 35 & 120 & 1 & 1 & \\
6 & 0 & 1 & 0 & 1 & 0 & 0 & 1 & 0 & 0 & 0 & 4 & 2 & 5 & 8 & 7 & 6 & 4 & 2 & 1 & 0 & 0 & 2 & 189 & 10 & 1 & 1 & \\
6 & 0 & 0 & 0 & 0 & 0 & 0 & 1 & 1 & 0 & 0 & 3 & 1 & 3 & 6 & 6 & 6 & 5 & 4 & 3 & 2 & 1 & 2 & 1 & 320 & 1 & 1 & $\textrm{top}$ \\
6 & 0 & 0 & 0 & 0 & 0 & 2 & 0 & 0 & 0 & 0 & 2 & 2 & 4 & 6 & 6 & 6 & 5 & 4 & 3 & 2 & 1 & 2 & 1 & 126 & 1 & 1 & $\textrm{top}$ \\
6 & 1 & 0 & 0 & 0 & 1 & 0 & 1 & 0 & 0 & 0 & 4 & 2 & 5 & 8 & 7 & 6 & 4 & 3 & 2 & 1 & 0 & 0 & 35 & 10 & 8 & 1 & \\
6 & 0 & 0 & 0 & 0 & 0 & 0 & 1 & 0 & 0 & 0 & 4 & 2 & 5 & 8 & 7 & 6 & 5 & 4 & 3 & 2 & 1 & -2 & 1 & 10 & 46 & 1 & $\textrm{top}$ \\
\hline
\end{longtable}

\begin{longtable}{|r|r@{\ }r@{\ }r@{\ }r|r@{\ }r@{\ }r@{\ }r@{\ }r@{\ }r|r@{\ }r@{\ }r@{\ }r@{\ }r@{\ }r@{\ }r@{\ }r@{\ }r@{\ }r@{\ }r|r|r|r|r|r|c|}
\caption{$A_4 \times E_6$ representations in $E_{11}$ ($D=5$)} \\
\hline
\multicolumn{1}{|c|}{$l$} &
\multicolumn{4}{|c|}{$p_{\rm{grav}}$} &
\multicolumn{6}{|c|}{$p_{\rm{int}}$} &
\multicolumn{11}{|c|}{$m$} &
\multicolumn{1}{|c|}{$\alpha^2$} &
\multicolumn{1}{|c|}{$d_{\rm{reg}}$} &
\multicolumn{1}{|c|}{$d_{\rm{int}}$} &
\multicolumn{1}{|c|}{$\textrm{mult}(\alpha)$} &
\multicolumn{1}{|c|}{$\mu$} & fields \\
\hline
\hline
0 & 0 & 0 & 0 & 0 & 1 & 0 & 0 & 0 & 0 & 0 & -2 & -1 & -2 & -3 & -2 & -1 & 0 & 0 & 0 & 0 & 0 & 2 & 1 & 78 & 1 & 1 & $p=0$ \\
0 & 1 & 0 & 0 & 1 & 0 & 0 & 0 & 0 & 0 & 0 & 0 & 0 & 0 & 0 & 0 & 0 & 0 & -1 & -1 & -1 & -1 & 2 & 24 & 1 & 1 & 1 & $\bar g_{\mu\nu}$ \\
0 & 0 & 0 & 0 & 0 & 0 & 0 & 0 & 0 & 0 & 0 & 0 & 0 & 0 & 0 & 0 & 0 & 0 & 0 & 0 & 0 & 0 & 0 & 1 & 1 & 11 & 1 & $\hat g_{\mu\nu}$ \\
\hline
1 & 1 & 0 & 0 & 0 & 0 & 0 & 0 & 0 & 0 & 1 & 0 & 0 & 0 & 0 & 0 & 0 & 1 & 0 & 0 & 0 & 0 & 2 & 5 & 27 & 1 & 1 & $p=1$ \\
\hline
2 & 0 & 1 & 0 & 0 & 0 & 1 & 0 & 0 & 0 & 0 & 1 & 0 & 1 & 2 & 2 & 2 & 2 & 1 & 0 & 0 & 0 & 2 & 10 & 27 & 1 & 1 & ${}^\star\,  (p=1)$ \\
\hline
3 & 0 & 0 & 1 & 0 & 1 & 0 & 0 & 0 & 0 & 0 & 1 & 1 & 2 & 3 & 3 & 3 & 3 & 2 & 1 & 0 & 0 & 2 & 10 & 78 & 1 & 1 & ${}^\star\,  (p=0)$ \\
3 & 1 & 1 & 0 & 0 & 0 & 0 & 0 & 0 & 0 & 0 & 3 & 2 & 4 & 6 & 5 & 4 & 3 & 1 & 0 & 0 & 0 & 2 & 40 & 1 & 1 & 1 & ${}^\star\,  g_{\mu\nu}$ \\
\hline
4 & 0 & 0 & 0 & 1 & 0 & 0 & 1 & 0 & 0 & 0 & 2 & 1 & 2 & 4 & 4 & 4 & 4 & 3 & 2 & 1 & 0 & 2 & 5 & 351 & 1 & 1 & $\textrm{de}$ \\
4 & 1 & 0 & 1 & 0 & 0 & 0 & 0 & 0 & 0 & 1 & 3 & 2 & 4 & 6 & 5 & 4 & 4 & 2 & 1 & 0 & 0 & 2 & 45 & 27 & 1 & 1 & \\
\hline
5 & 1 & 0 & 0 & 1 & 0 & 0 & 0 & 0 & 1 & 0 & 3 & 2 & 4 & 6 & 5 & 5 & 5 & 3 & 2 & 1 & 0 & 2 & 24 & 351 & 1 & 1 & \\
5 & 0 & 0 & 0 & 0 & 1 & 1 & 0 & 0 & 0 & 0 & 2 & 1 & 3 & 5 & 5 & 5 & 5 & 4 & 3 & 2 & 1 & 2 & 1 & 1728 & 1 & 1 & $\textrm{top}$ \\
5 & 0 & 1 & 1 & 0 & 0 & 1 & 0 & 0 & 0 & 0 & 4 & 2 & 5 & 8 & 7 & 6 & 5 & 3 & 1 & 0 & 0 & 2 & 75 & 27 & 1 & 1 & \\
5 & 1 & 0 & 0 & 1 & 0 & 1 & 0 & 0 & 0 & 0 & 4 & 2 & 5 & 8 & 7 & 6 & 5 & 3 & 2 & 1 & 0 & 0 & 24 & 27 & 8 & 1 & \\
5 & 0 & 0 & 0 & 0 & 0 & 1 & 0 & 0 & 0 & 0 & 4 & 2 & 5 & 8 & 7 & 6 & 5 & 4 & 3 & 2 & 1 & -2 & 1 & 27 & 46 & 1 & $\textrm{top}$ \\
\hline
\end{longtable}

\begin{longtable}{|r|r@{\ }r@{\ }r|r@{\ }r@{\ }r@{\ }r@{\ }r@{\ }r@{\ }r|r@{\ }r@{\ }r@{\ }r@{\ }r@{\ }r@{\ }r@{\ }r@{\ }r@{\ }r@{\ }r|r|r|r|r|r|c|}
\caption{$A_3 \times E_7$ representations in $E_{11}$ ($D=4$)} \\
\hline
\multicolumn{1}{|c|}{$l$} &
\multicolumn{3}{|c|}{$p_{\rm{grav}}$} &
\multicolumn{7}{|c|}{$p_{\rm{int}}$} &
\multicolumn{11}{|c|}{$m$} &
\multicolumn{1}{|c|}{$\alpha^2$} &
\multicolumn{1}{|c|}{$d_{\rm{reg}}$} &
\multicolumn{1}{|c|}{$d_{\rm{int}}$} &
\multicolumn{1}{|c|}{$\textrm{mult}(\alpha)$} &
\multicolumn{1}{|c|}{$\mu$} & fields \\
\hline
\hline
0 & 0 & 0 & 0 & 0 & 1 & 0 & 0 & 0 & 0 & 0 & -2 & -2 & -3 & -4 & -3 & -2 & -1 & 0 & 0 & 0 & 0 & 2 & 1 & 133 & 1 & 1 & $p=0$ \\
0 & 1 & 0 & 1 & 0 & 0 & 0 & 0 & 0 & 0 & 0 & 0 & 0 & 0 & 0 & 0 & 0 & 0 & 0 & -1 & -1 & -1 & 2 & 15 & 1 & 1 & 1 & $\bar g_{\mu\nu}$ \\
0 & 0 & 0 & 0 & 0 & 0 & 0 & 0 & 0 & 0 & 0 & 0 & 0 & 0 & 0 & 0 & 0 & 0 & 0 & 0 & 0 & 0 & 0 & 1 & 1 & 11 & 1 & $\hat g_{\mu\nu}$ \\
\hline
1 & 1 & 0 & 0 & 0 & 0 & 0 & 0 & 0 & 0 & 1 & 0 & 0 & 0 & 0 & 0 & 0 & 0 & 1 & 0 & 0 & 0 & 2 & 4 & 56 & 1 & 1 & $p=1,$ \\
 &  &  &  &  &  &  &  &  &  &  &  &  &  &  &  &  &  &  &  &  &  &  &  & &  &  & ${}^\star\, (p=1)$ \\
\hline
2 & 0 & 1 & 0 & 0 & 1 & 0 & 0 & 0 & 0 & 0 & 1 & 0 & 1 & 2 & 2 & 2 & 2 & 2 & 1 & 0 & 0 & 2 & 6 & 133 & 1 & 1 & ${}^\star\,  (p=0)$ \\
2 & 2 & 0 & 0 & 0 & 0 & 0 & 0 & 0 & 0 & 0 & 3 & 2 & 4 & 6 & 5 & 4 & 3 & 2 & 0 & 0 & 0 & 2 & 10 & 1 & 1 & 1 & ${}^\star\,  g_{\mu\nu}$ \\
\hline
3 & 0 & 0 & 1 & 1 & 0 & 0 & 0 & 0 & 0 & 0 & 1 & 1 & 2 & 3 & 3 & 3 & 3 & 3 & 2 & 1 & 0 & 2 & 4 & 912 & 1 & 1 & $\textrm{de}$ \\
3 & 1 & 1 & 0 & 0 & 0 & 0 & 0 & 0 & 0 & 1 & 3 & 2 & 4 & 6 & 5 & 4 & 3 & 3 & 1 & 0 & 0 & 2 & 20 & 56 & 1 & 1 & \\
\hline
4 & 0 & 0 & 0 & 0 & 0 & 1 & 0 & 0 & 0 & 0 & 2 & 1 & 2 & 4 & 4 & 4 & 4 & 4 & 3 & 2 & 1 & 2 & 1 & 8645 & 1 & 1 & $\textrm{top}$ \\
4 & 1 & 0 & 1 & 0 & 0 & 0 & 0 & 0 & 1 & 0 & 3 & 2 & 4 & 6 & 5 & 4 & 4 & 4 & 2 & 1 & 0 & 2 & 15 & 1539 & 1 & 1 & \\
4 & 0 & 2 & 0 & 0 & 1 & 0 & 0 & 0 & 0 & 0 & 4 & 2 & 5 & 8 & 7 & 6 & 5 & 4 & 2 & 0 & 0 & 2 & 20 & 133 & 1 & 1 & \\
4 & 1 & 0 & 1 & 0 & 1 & 0 & 0 & 0 & 0 & 0 & 4 & 2 & 5 & 8 & 7 & 6 & 5 & 4 & 2 & 1 & 0 & 0 & 15 & 133 & 8 & 1 & \\
4 & 0 & 0 & 0 & 0 & 1 & 0 & 0 & 0 & 0 & 0 & 4 & 2 & 5 & 8 & 7 & 6 & 5 & 4 & 3 & 2 & 1 & -2 & 1 & 133 & 46 & 1 & $\textrm{top}$ \\
4 & 2 & 1 & 0 & 0 & 0 & 0 & 0 & 0 & 0 & 0 & 6 & 4 & 8 & 12 & 10 & 8 & 6 & 4 & 1 & 0 & 0 & 2 & 45 & 1 & 1 & 1 & \\
4 & 1 & 0 & 1 & 0 & 0 & 0 & 0 & 0 & 0 & 0 & 6 & 4 & 8 & 12 & 10 & 8 & 6 & 4 & 2 & 1 & 0 & -2 & 15 & 1 & 44 & 1 & \\
\hline
\end{longtable}

\begin{longtable}{|r|r@{\ }r|r@{\ }r@{\ }r@{\ }r@{\ }r@{\ }r@{\ }r@{\ }r|r@{\ }r@{\ }r@{\ }r@{\ }r@{\ }r@{\ }r@{\ }r@{\ }r@{\ }r@{\ }r|r|r|r|r|r|c|}
\caption{$A_2 \times E_8$ representations in $E_{11}$ ($D=3$)} \\
\hline
\multicolumn{1}{|c|}{$l$} &
\multicolumn{2}{|c|}{$p_{\rm{grav}}$} &
\multicolumn{8}{|c|}{$p_{\rm{int}}$} &
\multicolumn{11}{|c|}{$m$} &
\multicolumn{1}{|c|}{$\alpha^2$} &
\multicolumn{1}{|c|}{$d_{\rm{reg}}$} &
\multicolumn{1}{|c|}{$d_{\rm{int}}$} &
\multicolumn{1}{|c|}{$\textrm{mult}(\alpha)$} &
\multicolumn{1}{|c|}{$\mu$} & fields \\
\hline
\hline
0 & 0 & 0 & 0 & 0 & 0 & 0 & 0 & 0 & 0 & 1 & -3 & -2 & -4 & -6 & -5 & -4 & -3 & -2 & 0 & 0 & 0 & 2 & 1 & 248 & 1 & 1 & $p=0$ \\
0 & 1 & 1 & 0 & 0 & 0 & 0 & 0 & 0 & 0 & 0 & 0 & 0 & 0 & 0 & 0 & 0 & 0 & 0 & 0 & -1 & -1 & 2 & 8 & 1 & 1 & 1 & $\bar g_{\mu\nu}$ \\
0 & 0 & 0 & 0 & 0 & 0 & 0 & 0 & 0 & 0 & 0 & 0 & 0 & 0 & 0 & 0 & 0 & 0 & 0 & 0 & 0 & 0 & 0 & 1 & 1 & 11 & 1 & $\hat g_{\mu\nu}$ \\
\hline
1 & 1 & 0 & 0 & 0 & 0 & 0 & 0 & 0 & 0 & 1 & 0 & 0 & 0 & 0 & 0 & 0 & 0 & 0 & 1 & 0 & 0 & 2 & 3 & 248 & 1 & 1 & ${}^\star\,  (p=0)$ \\
\hline
2 & 0 & 1 & 0 & 1 & 0 & 0 & 0 & 0 & 0 & 0 & 1 & 0 & 1 & 2 & 2 & 2 & 2 & 2 & 2 & 1 & 0 & 2 & 3 & 3875 & 1 & 1 & $\textrm{de}$ \\
2 & 2 & 0 & 0 & 0 & 0 & 0 & 0 & 0 & 0 & 1 & 3 & 2 & 4 & 6 & 5 & 4 & 3 & 2 & 2 & 0 & 0 & 2 & 6 & 248 & 1 & 1 & \\
2 & 0 & 1 & 0 & 0 & 0 & 0 & 0 & 0 & 0 & 0 & 6 & 4 & 8 & 12 & 10 & 8 & 6 & 4 & 2 & 1 & 0 & -2 & 3 & 1 & 44 & 1 & $\textrm{de}$ \\
\hline
3 & 0 & 0 & 1 & 0 & 0 & 0 & 0 & 0 & 0 & 0 & 1 & 1 & 2 & 3 & 3 & 3 & 3 & 3 & 3 & 2 & 1 & 2 & 1 & 147250 & 1 & 1 & $\textrm{top}$ \\
3 & 1 & 1 & 0 & 0 & 0 & 0 & 0 & 0 & 1 & 0 & 3 & 2 & 4 & 6 & 5 & 4 & 3 & 3 & 3 & 1 & 0 & 2 & 8 & 30380 & 1 & 1 & \\
3 & 1 & 1 & 0 & 1 & 0 & 0 & 0 & 0 & 0 & 0 & 4 & 2 & 5 & 8 & 7 & 6 & 5 & 4 & 3 & 1 & 0 & 0 & 8 & 3875 & 8 & 1 & \\
3 & 0 & 0 & 0 & 1 & 0 & 0 & 0 & 0 & 0 & 0 & 4 & 2 & 5 & 8 & 7 & 6 & 5 & 4 & 3 & 2 & 1 & -2 & 1 & 3875 & 46 & 1 & $\textrm{top}$ \\
3 & 3 & 0 & 0 & 0 & 0 & 0 & 0 & 0 & 0 & 1 & 6 & 4 & 8 & 12 & 10 & 8 & 6 & 4 & 3 & 0 & 0 & 2 & 10 & 248 & 1 & 1 & \\
3 & 1 & 1 & 0 & 0 & 0 & 0 & 0 & 0 & 0 & 1 & 6 & 4 & 8 & 12 & 10 & 8 & 6 & 4 & 3 & 1 & 0 & -2 & 8 & 248 & 44 & 1 & \\
3 & 0 & 0 & 0 & 0 & 0 & 0 & 0 & 0 & 0 & 1 & 6 & 4 & 8 & 12 & 10 & 8 & 6 & 4 & 3 & 2 & 1 & -4 & 1 & 248 & 206 & 1 & $\textrm{top}$ \\
3 & 1 & 1 & 0 & 0 & 0 & 0 & 0 & 0 & 0 & 0 & 9 & 6 & 12 & 18 & 15 & 12 & 9 & 6 & 3 & 1 & 0 & -4 & 8 & 1 & 192 & 1 & \\
\hline
\end{longtable}

\end{document}